\documentclass[aps,showpacs,superscriptaddress,prd]{revtex4}
\usepackage{amssymb,amsmath}

\begin{document}
\title{Dynamical renormalization group approach to the Altarelli-Parisi-Lipatov equations}
\author{D. Boyanovsky}
\affiliation{Department of Physics and Astronomy, University of
Pittsburgh, Pittsburgh, Pennsylvania 15260, USA}
\author{H. J. de Vega}
\affiliation{LPTHE, Universit\'e Pierre et Marie Curie (Paris VI)
et Denis Diderot (Paris VII), Tour 16, 1er. \'etage, 4, Place
Jussieu, 75252 Paris, Cedex 05, France}
\author{D.-S. Lee}
\affiliation{Department of Physics, National Dong Hwa University,
Shoufeng, Hualien 974, Taiwan, ROC}
\author{S.-Y. Wang}
\affiliation{Department of Physics and Astronomy, University of
Pittsburgh, Pittsburgh, Pennsylvania 15260, USA}
\author{H.-L. Yu}
\affiliation{Institute of Physics, Academia Sinica, Taipei, Taiwan
115, ROC}
\date{\today}

\begin{abstract}
The Altarelli-Parisi-Lipatov equations for the parton distribution
functions are rederived using the dynamical renormalization group
approach to quantum kinetics. This method systematically treats
the $\ln Q^2$ corrections that arises in perturbation theory as a
renormalization of the parton distribution function and
unambiguously indicates that the strong coupling must be allowed
to run with the scale in the evolution kernel. To leading
logarithmic accuracy the evolution equation is Markovian and the
logarithmic divergences in the perturbative expansion are
identified with the secular divergences (terms that grow in time)
that emerge in a perturbative treatment of the kinetic equations
in nonequilibrium systems. The resummation of the leading
logarithms by the Altarelli-Parisi-Lipatov equation is thus
similar to the resummation of the leading secular terms by the
Boltzmann kinetic equation.
\end{abstract}


\maketitle

An early important test of QCD and one that validates the parton
picture, is the $Q^2$ evolution of the quark distribution, or
structure function, in deep inelastic lepton-hadron scattering
which reveals scaling violations as predicted by QCD. The leading
logarithmic dependence of the quark structure function on the
scale $Q^2$ in deeply inelastic scattering was originally computed
in QCD via the operator product expansion~\cite{OPE}. In
perturbation theory, parton emission leads to QCD corrections to
the distribution functions that grow as $\ln Q^2$, which
eventually become large and invalidate the perturbative expansion.
These corrections signal the violations of scaling in deep
inelastic scattering.

Lipatov~\cite{lipatov} and Altarelli and Parisi~\cite{AP} provided
an alternative derivation of the evolution equations for the
parton distribution functions by associating the $\ln Q^2$
corrections with a \emph{change} in the parton distribution
function at the scale $Q^2$. These authors provided a set of
integro-differential equations that resum the perturbative
expansion and describe the evolution of the parton densities as a
function of $\ln Q^2$~\cite{peskin}. These parton evolution
equations are successful in predicting the correct leading
logarithmic dependence of the distribution functions in $Q^2$ and
the scaling violations predicted by these equations have been
confirmed spectacularly in deep inelastic
experiments~\cite{handbook}.

Altarelli and Parisi used a probabilistic description in their
derivation, however infrared singularities associated with the
emission of low momentum partons introduce singularities in some
probability densities. Anticipating that the infrared
singularities should cancel between real and virtual soft gluons,
these authors provided a prescription to handle these
singularities.

Collins and Qiu~\cite{CQ} provided a derivation of the
Altarelli-Parisi-Lipatov (APL) equations with a special treatment
of virtual diagrams that clearly reveals the explicit cancellation
of the infrared singularities between the contributions of real
and virtual soft gluons through Ward identities. The result of
Collins and Qiu is contained in a set of integro-differential
equations, equivalent to the APL ones but without the need of
extra prescriptions for handling soft gluon emission since the
cancellation of infrared singularities is manifest in their
treatment of real and virtual diagrams. For valence quarks these
equations have the form
\begin{equation}
Q^2\frac{ d}{dQ^2}\,q_i(x,Q^2) =
\frac{\alpha_s(Q^2)}{2\pi}\left[\int^1_{x}
\frac{dx_1}{x_1}\,q_i(x_1,Q^2)\,\gamma_{qq}\left(\frac{x}{x_1}\right)-
q_i(x,Q^2)\int^1_{0}dz\,\gamma_{qq}(z)\right], \label{eqCQ}
\end{equation}
where $q_i(x,Q^2)$ is the distribution function of the (valence)
$i$-quark at a fixed momentum fraction $x$ that evolves with the
momentum scale $Q^2$ at which the distribution function is probed
and the quark-quark splitting function is given by
\begin{equation}\label{splitfun}
\gamma_{qq}(z)=\frac{N^2_c -
1}{2N_c}\left(\frac{1+z^2}{1-z}\right)
\end{equation}
with $N_c$ the number of colors. The splitting function
$\gamma_{qq}(z)$ describe the probability density of finding a
parton quark inside another parton quark with momentum fraction
$z$ of the parent parton momentum and was originally introduced by
Weiszacker and Williams for QED~\cite{WW}. The relative minus sign
between the two terms allows an explicit cancellation of the
infrared divergences arising from (real and virtual) soft gluon
emission~\cite{CQ}.

A remarkable aspect of the Altarelli-Parisi-Lipatov equations in
the form obtained by Collins and Qiu given by Eq.~(\ref{eqCQ})
above, is that they feature exactly the same splitting function in
both terms on the right-hand side. This feature allows an
unambiguous probabilistic interpretation as a set of
\emph{Boltzmann kinetic equations}, since the different
contributions can be interpreted as ``gain'' and ``loss''
processes that change the parton distribution in ``cells''
labelled by the variable $x$, the fractional momentum of the
hadron carried by the parton~\cite{CQ,DP}.

Within the context of the probabilistic interpretation suggested
by Collins and Qiu and for reasons that will become clear below,
it is important to remark at this stage that the
Altarelli-Parisi-Lipatov equation involves a convolution in the
variable $x$ but \emph{not} in the variable $Q^2$, i.e., the
parton distribution function on the right-hand side of the
equation is evaluated at the same $Q^2$ as that on the left-hand
side. This feature of the evolution equation results in a
\emph{Markovian} approximation of parton evolution inside a
hadron. Physically, this implies that the evolution of the parton
distribution at some fixed $x$ involves only parton emission or
absorption processes taking place at \emph{momentum scale} $Q^2$
where the parton distribution is probed, rather than those taking
place at any other momentum scales. This property is actually
inherent from the leading logarithmic approximation employed in
the formulation, where each parton emission or absorption process
is actually independent of the others and depends only on the
momentum scale at which the process takes place.

This interpretation of the Altarelli-Parisi-Lipatov equations as a
set of kinetic equations for the parton distribution functions
motivates us to provide yet another alternative derivation of
these equations by implementing a novel approach to quantum
kinetics introduced recently in the form of the \emph{dynamical
renormalization group}~\cite{boyanrgkin}.

While we anticipate that the approach presented here will
ultimately lead to the form of the Altarelli-Parisi-Lipatov
equations derived by Collins and Qiu, we believe that this
alternative formulation yields novel insights that could prove
valuable to implement a systematic treatment of parton evolution.
As will be shown in detail below, the rederivation of the
evolution equations via the dynamical renormalization group
provides further and deeper understanding of many important
aspects:
\begin{itemize}
\item[(i)]{It reveals some subtle aspects of the
derivation which highlight the role of the running of the strong
coupling and the ``separation of scales'' which is a necessary
ingredient in a kinetic description. The role of the running
coupling is ultimately at the heart of the scaling violations and
the necessity for including the running of the coupling in the
evolution kernel becomes manifest in this approach. }
\item[(ii)]{The dynamical renormalization
group directly reveals how the $\ln Q^2$ that arise in the
perturbative expansion can be systematically absorbed into the
parton distribution functions and establishes a direct
correspondence with \emph{secular divergences in
time}~\cite{boyanrgkin} that are manifest in the derivation of the
Boltzmann equation and also responsible for the failure of the
perturbative expansion in quantum kinetics.}
\item[(iii)]{It directly establishes a correspondence between
the leading logarithmic approximation to parton evolution equation
which is local in $Q^2$ and the Markovian approximation to kinetic
equation which is local in time.}
\end{itemize}

As noticed by Collins and Qiu and mentioned above, the connection
between the Altarelli-Parisi-Lipatov equation and the kinetic
equation in nonequilibrium statistical physics is established by
identifying $\ln Q^2$ with the ``time'' variable. This
identification is the main concept upon which the formulation
based on the dynamical renormalization group hinges. In this
article, we revisit the derivation of Eq.~(\ref{eqCQ}) but
implementing the concepts of the dynamical renormalization
group~\cite{boyanrgkin}.

The $i$-quark distribution function at some scale $Q^2$ with fixed
momentum fraction $x$ is defined as~\cite{CQ}
\begin{equation}
q_i(x,Q^2)=(2\pi)^3 2E_p\int\frac{d^4p}{2P\cdot n}\,
T^i_{ba}(P,p)\,(\gamma_-)_{ba}\,\delta\left(x-\frac{p\cdot
n}{P\cdot n}\right), \label{pdf}
\end{equation}
with
\begin{equation}
T^i_{ba}(P,p)=\frac1{(2\pi)^4}\int d^4y\, e^{-ip\cdot y} \langle
P|\overline{\psi}_b(y) \psi_a(0)| P\rangle,
\end{equation}
where $n_\mu = (n_+, n_-,n_\perp)=(1, 0, 0_\perp)$ and $P$ is the
momentum of the hadron. The derivation of the parton evolution
equations to a given order in the strong coupling, begins by
expanding $T_{ba}^i$ in perturbation theory in $\alpha_s$, and
evaluating  the corresponding Feynman diagrams at large $Q^2$ and
$P$ with $x=Q^2/2P\cdot{q}$ fixed.

The main idea followed in Ref.~\onlinecite{CQ} is to calculate the
\emph{change} in the parton distribution function from gluon
radiation processes. In the infinite momentum frame as the scale
$Q^2$ changes the number of partons with momentum fraction $x$
also changes. Consider the number of partons with momentum
fraction $x$ at the scale $Q^2$. On the one hand, partons with
larger momentum fraction $x_1>x$ at the scale $Q^2$ that emit a
gluon with momentum fraction $x_1-x$ now enter into the ``phase
space cell'' with fractional momentum $x$. On the other hand,
partons that are already in the ``phase space cell'' with momentum
fraction $x$ at the scale $Q^2$ leave this ``cell'' by diminishing
their value of $x$ upon radiating a gluon. The former corresponds
to the ``gain'' contribution, whereas the latter corresponds to
the ``loss'' contribution.

Collins and Qiu showed that the real gluon emission processes
gives a positive contribution to the parton distribution function
and can be calculated from the real gluon diagram by introducing a
cutoff $\Lambda^2$~\cite{CQ}. Most importantly, by converting the
virtual gluon diagrams into a real diagram which has the same form
as that corresponding to real gluon emission except for having the
opposite sign, they obtained an explicit cancelation of infrared
divergences between diagrams and the ``gain'' and ``loss''
contributions to the parton distribution function~\cite{CQ}
\begin{equation}
\Delta q_i(x,\Lambda^2)= \Delta q_i(x,\Lambda^2)_\mathrm{gain}-
\Delta q_i(x,\Lambda^2)_\mathrm{loss}, \label{eq3}
\end{equation}
with
\begin{eqnarray}
\Delta q_i(x,\Lambda^2)_\mathrm{gain} & = & \int^{\Lambda^2}
\frac{dk^2_\perp}{2\pi k^2_\perp}\,\alpha_s(k^2_{\perp}) \int^1_x
\frac{dx_1}{x_1}\,q_i(x_1,k_\perp^2)\,
\gamma_{qq}\left(\frac{x}{x_1}\right), \nonumber\\
\Delta q_i(x,\Lambda^2)_\mathrm{loss} & = &\int^{\Lambda^2}
\frac{dk^2_\perp}{2\pi
k^2_\perp}\,\alpha_s(k^2_{\perp})\,q_i(x,k_\perp^2)\int^1_0 dz \,
\gamma_{qq}(z),\label{gainloss}
\end{eqnarray}
where the quark-quark splitting function $\gamma_{qq}(z)$ is given
by Eq.~(\ref{splitfun}). Taking the derivative with respect to
$\ln\Lambda^2$ and identifying the cutoff scale $\Lambda^2$ as the
scale $Q^2$ where the parton distribution function is
probed~\cite{CQ}, Collins and Qiu obtained the parton evolution
equation for valence quarks as given by Eq.~(\ref{eqCQ}).

A close examination the above equations reveal the following
important features in their derivation:
\begin{itemize}
\item[(i)]{The quark distribution function that enters the
right-hand side of the gain and loss contributions given by
Eq.~(\ref{gainloss}) carries the $k_\perp^2$ dependence. Since
$k_\perp^2$ is integrated out to give the total change $\Delta
q_i(x,\Lambda^2)$, this in turn is tantamount to an implicit
resummation of the leading logarithms  in Collins and Qiu's
derivation~\cite{CQ} (see also below).}
\item[(ii)]{The strong coupling $\alpha_s$ is taken to be \emph{running} with the
scale of the exchanged gluon in the process. While the lowest
order calculation of the relevant Feynman diagram given by Fig.~5
in Ref.~\onlinecite{CQ} would be in terms of the bare gluon
propagator and the bare strong coupling constant, in allowing the
strong coupling to run with the scale, the exchanged gluon has
been effectively ``dressed'' by parton loops.}
\end{itemize}

We address these issues in detail by implementing the novel
dynamical renormalization group approach to quantum
kinetics~\cite{boyanrgkin}. This approach  reveals in an
alternative manner, the type of resummations implied by the final
form of the parton evolution equation. This alternative
formulation is based on Collins and Qiu's probabilistic
interpretation of the parton evolution equation as a Boltzmann
kinetic equation with $\ln Q^2$ interpreted as a ``time'' variable
and $x$ as a ``momentum'' or ``cell'' variable in phase space. In
particular, as will be shown below, the dynamical renormalization
group approach to the Altarelli-Parisi-Lipatov equations not only
provides and justifies an alternative and explicit resummation of
the leading logarithms in the form a resummation of the leading
secular terms, but also indicates unambiguously the running of the
strong coupling in Collins and Qiu's derivation is crucial to
violations of scaling.

The derivation of the kinetic equation for the
\emph{single-particle distribution functions} in phase space via
the dynamical renormalization group~\cite{boyanrgkin} begins by
considering that the distribution function is determined at some
initial time $t_0$ at which the initial state of the
nonequilibrium system is specified. That is, the initial condition
for the time evolution is determined by the initial distribution
function $n_\mathbf{k}(t_0)$ for particles of momentum
$\mathbf{k}$. Time evolution through the interaction, changes the
number of particles in a given phase space cell, say, labeled by
momentum $\mathbf{k}$. The evolved distribution function at a
later time $t=t_0+\Delta t$ is calculated in perturbation theory
and can be written as $n_\mathbf{k}(t) = n_\mathbf{k}(t_0)+\Delta
n_\mathbf{k}[n_\mathbf{k}(t_0),\Delta t]$. The change in the
distribution function depends on the distribution function at the
initial time $n_\mathbf{k}(t_0)$ because the perturbative
expansion is carried out with respect to the initial
state~\cite{boyanrgkin}; the dependence on $\Delta t$ is
determined by the transition probability which is obtained by
using time-dependent perturbation theory. For $\Delta t$ much
larger than any microscopic time scales, using Fermi's golden rule
one finds that the transition probability is proportional to
$\Delta t$ so that $\Delta n_\mathbf{k}[n_\mathbf{k}(t_0),\Delta
t]= C[n_\mathbf{k}(t_0)]\,\Delta t$, where the collision term
$C[n_\mathbf{k}(t_0)]$ depends on the distribution functions at
$t_0$. Hence, one obtains
\begin{equation}
\frac{\Delta n_\mathbf{k}(t)}{\Delta t}= C[n_\mathbf{k}(t_0)],
\label{albol}
\end{equation}
which is \emph{almost} the form of the Boltzmann equation, were it
not for the fact that the term on the right-hand side depends on
the distribution function at the initial time. This in turn
implies that the change in the distribution function at long times
is proportional to $\Delta t$. Thus perturbation theory features
\emph{secular terms}, i.e., terms that grow in time and would lead
to the breakdown of the perturbative expansion.

The Boltzmann equation is obtained after a \emph{resummation}
directly in real time via the dynamical renormalization
group~\cite{boyanrgkin}. The resummation implied by the dynamical
renormalization group can be summarized as follows: Consider
evolving in time from $t_0$ to $t_0+\Delta t$ with $\Delta t$ much
larger than a typical microscopic time so that Fermi's golden rule
can be applied, but not too long so that perturbation theory is
still valid during this time interval. This is exactly the
assumption of the separation of time scales in a kinetic
description. At the end of this interval the new distribution
function is calculated as $n_\mathbf{k}(t_0+\Delta
t)=n_\mathbf{k}(t_0)+C[n_\mathbf{k}(t_0)]\,\Delta t$. Now use this
new distribution function as the initial value at $t_0+\Delta t$
to evolve further another time step, i.e., the distribution
function is ``reset'' to the new value after each evolution in a
time step. This resummation procedure can be implemented
consistently via the dynamical renormalization
group~\cite{boyanrgkin} and leads to the Boltzmann kinetic
equation
\begin{equation}
\frac{d}{dt}n_\mathbf{k}(t)= C[n_\mathbf{k}(t)], \label{corrboltz}
\end{equation}
whose solution is found to be given by
\begin{equation} n_\mathbf{k}(t)=n_\mathbf{k}(t_0)+\Delta
n_\mathbf{k}(t),\quad \Delta n_\mathbf{k}(t)=\int^t_{t_0}
dt'\,C[n_\mathbf{k}(t')]. \label{boltzsol}
\end{equation}
A comparison of Eq.~(\ref{boltzsol}) and the gain and loss
contributions to the change of the parton distribution given by
Eq.~(\ref{gainloss}) reveals clearly that a resummation is
implicitly invoked in Collins and Qiu's derivation.

We now implement a resummation via the dynamical renormalization
group to obtain the evolution equation for the parton distribution
function in the same manner. Following Ref.~\onlinecite{CQ} we now
focus on the evolution equations for the distribution function of
valence quarks resulting from gluon emission. Those for the sea
quarks can be obtained by a straightforward generalization once
the relevant Feynman diagrams have been computed.

Consider that the parton distribution $q_i(x,Q^2_0)$ at a scale
$Q^2_0$ is given and that the evolution from the scale $Q^2_0$ to
a larger scale $Q^2$ results in a change of the parton
distribution as a consequence of parton radiation. These changes
are calculated via a perturbative expansion in the strong coupling
constant $\alpha_s$ by following exactly the steps summarized
above and described in detail in Ref.~\onlinecite{CQ}. The freedom
to choose the factorization and renormalization scale permits to
identify the upper momentum cutoff $\Lambda^2$ as $Q^2$ in the
loop integrals leading to $\Delta q_i(x,Q^2)$~\cite{CQ}. Taking
the lower limit to be the scale $Q^2_0$ and the distribution
function that enters in the diagrams at the scale $Q^2_0$
consistent with the leading logarithmic approximation, we obtain
\begin{eqnarray}
&&q_i(x,Q^2) = q_i(x,Q^2_0)+\Delta q_i(x,Q^2;Q^2_0),\nonumber\\
&&\Delta q_i(x,Q^2;Q^2_0)\equiv\Delta
q_i(x,Q^2;Q^2_0)_\mathrm{gain}- \Delta
q_i(x,Q^2;Q^2_0)_\mathrm{loss},\label{neweq3}
\end{eqnarray}
where
\begin{eqnarray}
&&\Delta q_i(x,Q^2;Q^2_0)_\mathrm{gain}  =  \int^{Q^2}_{Q^2_0}
\frac{dk^2_\perp}{2\pi k^2_\perp}\,\alpha_s(k^2_{\perp}) \int^1_x
\frac{dx_1}{x_1}\,q_i(x_1,Q^2_0)\,
\gamma_{qq}\left(\frac{x}{x_1}\right),\nonumber\\
&&\Delta q_i(x,Q^2;Q^2_0)_\mathrm{loss} = \int^{Q^2}_{Q^2_0}
\frac{dk^2_\perp}{2\pi
k^2_\perp}\,\alpha_s(k^2_{\perp})\,q_i(x,Q^2_0)\int^1_0 dz \,
\gamma_{qq}(z), \label{newgainloss}\\
&&\alpha_s(k^2_\perp)=\frac{\alpha_0}{\ln(k^2_\perp/\Lambda_\mathrm{QCD}^2)},\quad
\alpha_0=\frac{12\pi}{33-2N_f}, \nonumber
\end{eqnarray}
and $N_f$ is the number of flavors.

We emphasize an important aspect of the above equations that is
\emph{different} from those obtained by Collins and Qiu: the
parton distribution function enters at the \emph{initial scale}
$Q^2_0$. Thus in the ``update'' equation (\ref{neweq3}) with the
gain and loss contributions given by Eq.~(\ref{newgainloss}) the
parton distribution function is only convoluted in the variable
$x$ but not in $k^2_\perp$. This aspect in turn allows to obtain
the characteristic scales of parton evolution already at the
perturbative level. The dependence of the updated parton evolution
on the \emph{initial} parton distribution $q_i(x,Q^2_0)$ is
reminiscent of the derivation of the Boltzmann kinetic equation
from the time evolution of a given \emph{initial} state of a
nonequilibrium system outlined above.

Although, following Collins and Qiu, we have included the running
of the strong coupling in the gain and loss terms, we now study in
detail the justification for this choice. Consider first the above
expression in strict perturbation theory by taking the strong
coupling to be a constant $\alpha_s(Q_0)$ fixed at the initial
scale $Q_0$. This corresponds to considering a bare exchanged
gluon in the corresponding Feynman diagram. Performing the
$k^2_\perp$ integral, one obtains
\begin{equation}
\Delta q_i(x,Q^2;Q^2_0) \propto \alpha_s(Q^2_0)\, \ln(Q^2/Q^2_0).
\label{seclog}
\end{equation}
In the interpretation of $\ln Q^2$ as the ``time'' variable $t$,
this in turn translates into the statement
\begin{equation}
\Delta q_i(x,t;t_0) \propto \alpha_s(t_0)\,\left(t-t_0 \right),
\label{sect}
\end{equation}
which is akin to the expression for the \emph{real-time} evolution
of a single-particle distribution function calculated in strict
perturbation theory by using Fermi's golden rule as described
above.

In the dynamical renormalization group
approach~\cite{boyanrgkin,DRG} these type of terms (divergences in
the long-time limit) are identified as \emph{secular divergences}
in the perturbative expansion which signal the breakdown of
perturbation theory and can be resummed systematically via the
dynamical renormalization group.

Equation (\ref{seclog}) reveals that the parton distribution
function evolves on ``time'' scales (not to be confused with the
real time) \emph{comparable} to the scale of the running of the
strong coupling constant. Therefore during the ``time'' scale of
evolution of the parton distribution the coupling actually changes
substantially and its ``time'' evolution cannot be neglected. This
is to say that keeping the strong coupling as a constant
(determined at the initial scale) does \emph{not} allow a clear
separation of ``time'' scales between the ``kinetic'' scales
characteristic to parton evolution and the ``microscopic'' scale
determined by the strong coupling since they are of the same order
in the logarithmic approximation.

Including the running of the strong coupling in the loop momentum
integral is tantamount to including the vacuum polarization in the
exchanged gluon and leads to
\begin{equation}
\Delta q_i(x,Q^2;Q^2_0)\propto \alpha_0
\ln\left[\frac{\ln(Q^2/\Lambda_\mathrm{QCD}^2)}{\ln(Q^2_0/\Lambda_\mathrm{QCD}^2)}\right],
\label{secloglog}
\end{equation}
or, in terms of the ``time'' variable $t$,
\begin{equation}
\Delta q_i(x,t;t_0)\propto \alpha_0 \ln(t/t_0). \label{secloglogt}
\end{equation}
Therefore, including the running of the coupling in the evolution
kernel leads to a \emph{slower} evolution of the parton
distribution function. This is clearly a consequence of the fact
that during the ``time'' scale of evolution of the distribution
function the strong coupling changes substantially, becoming
smaller and therefore in turn slowing down the change in the
parton distribution.

This is the important issue of ``time'' scales that emerges in the
kinetic description. If the coupling varies on ``time'' scales
\emph{longer} than that of parton evolution, then it can be kept
constant during the evolution since there is a wide separation of
``time'' scales. However, if the coupling varies substantially
during the ``time'' scale of parton evolution, then the running of
the coupling must be included in the ``update'' equation of the
parton distribution function. Hence, this formulation based on the
``update'' equation (\ref{neweq3}) with the gain and loss
contributions given by Eq.~(\ref{newgainloss}) in terms of the
parton distribution function at the initial scales allows to
clarify unambiguously whether or not to include the running of the
strong coupling by comparing the ``time'' scale of parton
evolution and that of the running coupling. The criterion is as
follows: If the scales are comparable or the parton distribution
evolves \emph{slower}, then the running of the coupling must be
included; if the parton evolution evolves much faster, then the
coupling can be kept fixed.

In the case of evolution of the parton distribution functions in
QCD, the running of the strong coupling must be included in the
evolution kernel of the ``update'' equation. A similar situation
in which the evolution kernel depends explicitly on time has been
found in studying quantum kinetic of charged particles in a hot
QED (QCD) plasma as a result of infrared divergences associated
with the emission and absorption of quasistatic, soft magnetic
photons (gluons)~\cite{blaizot,boyanrgkin}. The similarity between
these cases is tantalizing.

Furthermore, the above ``update'' equation (\ref{neweq3}) with the
gain and loss contributions given by Eq.~(\ref{newgainloss})
implies that for fixed $x$ the change in the parton distribution
function due to a change in momentum scale only depends on the
parton distribution at the initial scale $Q_0^2$, i.e., the
momentum scale at which the gluon is emitted, rather than on the
parton distribution at intermediate scales between $Q_0^2$ and
$Q^2$. We want to stress once again that this result is consistent
with the leading logarithmic approximation where gluon emission at
each vertex is independent of the others and depends only on the
momentum scale at which the gluon is emitted. It is this feature
of the ``update'' equation that establishes a correspondence
between the leading logarithmic approximation in parton evolution
and the Markovian approximation in a kinetic
description~\cite{balescu}.

The logarithmic divergences for large $Q^2$ either in the form of
Eq.~(\ref{seclog}) or (\ref{secloglog}) clearly signals the
breakdown of the perturbative expansion and the necessity for
resummation. These divergences are akin to the secular divergences
that appear in time-dependent perturbation theory, in the
derivation of kinetic equations~\cite{boyanrgkin}, as well as in
the asymptotic analysis of differential equations~\cite{DRG}. A
systematic method based on the idea of renormalization group has
been developed in Ref.~\onlinecite{DRG} to obtain asymptotically
convergent solutions of differential equations whose perturbative
solution features secular terms. This renormalization group
method, which we now follow, has been recently generalized to the
realm of nonequilibrium quantum field theory and quantum
kinetics~\cite{boyanrgkin}.

The main idea of the renormalization group method, as described
above in obtaining the Boltzmann equation, is to evolve the parton
distribution function from an initial scale $Q^2_0$ in
perturbation theory up to a larger scale $Q^2_0+\Delta Q^2$. The
change in scale $\Delta Q^2$ is small enough so that the change in
the parton distribution is perturbatively small and hence the
perturbative expansion is reliable. The change in the parton
distribution is absorbed into the updated parton distribution at
the scale $Q^2_0+\Delta Q^2$, which is then used for the next
evolution in scale, and so on.
This procedure is implemented systematically by introducing a
``renormalization constant'' of the \emph{initial} parton
distribution ${\cal{Z}}(x,\tau^2,Q^2_0)$ at an intermediate but
otherwise arbitrary scale $\tau^2$, i.e.,
\begin{gather}
q_i(x,Q^2_0)={\cal{Z}}(x,\tau^2,Q^2_0)\,q_i(x,\tau^2),\nonumber\\
{\cal{Z}}(x,\tau^2,Q^2_0) =  1 +
z^{(1)}(x,\tau^2,Q^2_0)+z^{(2)}(x,\tau^2,Q^2_0)+\cdots \label{eq7}
\end{gather}
where $q_i(x,\tau^2)$ is the ``renormalized'' parton distribution
at scale $\tau^2$ and
$z^{(1)}(x,\tau^2,Q^2_0)\sim\mathcal{O}(\alpha_s)$,
$z^{(2)}(x,\tau^2,Q^2_0)\sim\mathcal{O}(\alpha^2_s)$, etc.
Substituting Eq.~(\ref{eq7}) into Eq.~(\ref{neweq3}) and keeping
terms only up to ${\cal{O}}(\alpha_s)$, one obtains
\begin{eqnarray}
q_i(x,Q^2)&=&q_i(x,\tau^2)[1 +z^{(1)}(x,\tau^2,Q^2_0)]+
\int^{Q^2}_{Q^2_0} \frac{dk^2_{\perp}}{2\pi
k^2_\perp}\,\alpha_s(k^2_{\perp}) \int^1_x
\frac{dx_1}{x_1}\,q_i(x_1,\tau^2) \nonumber\\
&& -\int^{Q^2}_{Q^2_0} \frac{dk^2_\perp}{2\pi
k^2_\perp}\,\alpha_s(k^2_{\perp})\,q_i(x,\tau^2)\int^1_0 dz\,
\gamma_{qq}(z) + {\cal{O}}(\alpha^2_s). \label{renoAP}
\end{eqnarray}
The coefficients $z^{(i)}(x,\tau^2,Q^2_0)$ are chosen to cancel
exactly the secular (logarithmic) terms at the scale $\tau$ order
by order in the perturbative expansion. Thus, to
${\cal{O}}(\alpha_s)$, one obtains
\begin{eqnarray}
 z^{(1)}(x,\tau^2,Q^2_0)&=&  -\frac{1}{q_i(x,\tau^2)}\left[
\int^{\tau^2}_{Q^2_0} \frac{dk^2_\perp}{2\pi k^2_\perp}\,
\alpha_s(k_\perp^2) \int^1_x \frac{dx_1}{x_1}\,q_i(x_1,\tau^2)\,
\gamma_{qq}\left(\frac{x}{x_1}\right)\right.\nonumber
\\ && - \left. \int^{\tau^2}_{Q^2_0} \frac{dk^2_\perp}{2\pi
k^2_\perp}\,\alpha_s(k^2_{\perp})\,q_i(x,\tau^2)\int^1_0 dz\,
\gamma_{qq}(z) \right]. \label{eq9}
\end{eqnarray}
Upon substituting this choice for $z^{(1)}(x,\tau^2,Q^2_0)$ back
into Eq.~(\ref{renoAP}), we arrive at
\begin{eqnarray}
q_i(x,Q^2)&=& q_i(x,\tau^2)+  \int^{Q^2}_{\tau^2}
\frac{dk^2_\perp}{2\pi k^2_\perp}\,\alpha_s(k_\perp^2) \int^1_x
\frac{dx_1}{x_1}\,q_i(x_1,\tau^2)\,
\gamma_{qq}\left(\frac{x}{x_1}\right)\nonumber\\
&&-\int^{Q^2}_{\tau^2} \frac{dk^2_\perp}{2\pi
k^2_\perp}\,\alpha_s(k^2_{\perp})\,q_i(x,\tau^2)\int^1_0 dz\,
\gamma_{qq}(z) + {\cal O}(\alpha^2_s), \label{eq10}
\end{eqnarray}
which can be improved by choosing the arbitrary renormalization
scale $\tau^2$ close to the scale $Q^2$. This renormalization
procedure embodies the original idea behind the
Altarelli-Parisi-Lipatov equation that the change in the parton
distribution due to a change in the scale can be absorbed into a
redefinition, indeed a renormalization, of the parton
distribution.

Since the scale $\tau^2$ is arbitrary, a change in the
renormalization scale is compensated by the corresponding change
in the ``renormalized'' parton distribution so that $q_i(x,Q^2)$
is independent of the renormalization scale $\tau^2$. This in turn
leads to the \emph{dynamical renormalization group equation}
$\tau^2 d q_i(x,Q^2)/d\tau^2 =0$, which to lowest order in
$\alpha_s$ is given by
\begin{equation}
\tau^2\frac{d}{d\tau^2}\,q_i(x,\tau^2)=
\frac{\alpha_s(\tau^2)}{2\pi} \left[ \int^1_x
\frac{dx_1}{x_1}\,q_i(x_1,\tau^2)\,
\gamma_{qq}\left(\frac{x}{x_1}\right)- \,q_i(x,\tau^2)\int^1_0
dz\, \gamma_{qq}(z)\right]. \label{eq12}
\end{equation}
Upon identifying the renormalization scale $\tau^2$ as the scale
$Q^2$, one finds immediately that Eq.~(\ref{eq12}) is exactly the
Altarelli-Parisi-Lipatov equation derived by Collins and Qiu.
Consequently, we find that while the dynamical renormalization
group leads to a kinetic equation that resums the \emph{leading
secular terms} in the perturbative expansion in nonequilibrium
quantum statistical mechanics, the Altarelli-Parisi-Lipatov
equation resums the \emph{leading logarithms}, which as argued
above are identified with the secular terms in a kinetic
description.

The solutions of Eq.~(\ref{eq12}) can be found by taking moments
of the parton distribution and quark-quark splitting
functions~\cite{CQ}
\begin{gather}
F^n_i(Q^2) = \int^1_0 dx\,x^{n-1}\,
q_i(x,Q^2), \nonumber\\
\gamma^n_{qq} = \int^1_0 dz\,(z^{n-1}-1)\,\gamma_{qq}(z).
\label{moments}
\end{gather}
After some algebra, one finds
\begin{equation}
Q^2\frac{d}{dQ^2}F^n_i(Q^2)=
-\frac{\alpha_s(Q^2)}{2\pi}\,\gamma^n_{qq}\, F^n_i(Q^2),
\end{equation}
where
\begin{equation}
\gamma^n_{qq} = \frac{N^2_c - 1}{2N_c}\left[
\frac{1}{2}-\frac{1}{n(n+1)}+2\sum_{l=2}^n\frac{1}{l}\right].
\label{diffeq}
\end{equation}
With the running strong coupling given by Eq.~(\ref{newgainloss}),
the solution is found to be given by
\begin{equation}
F^n_i(Q^2) = F^n_i(Q^2_0) \left[ \frac{
\ln(Q^2/\Lambda^2_\mathrm{QCD})}{\ln(Q^2_0/\Lambda^2_\mathrm{QCD})}
\right]^{-\alpha_0 \gamma^n_{qq}/2\pi}.\label{scal}
\end{equation}
The mild violation of scaling, a logarithm raised to a power,
instead of a pure power law with an anomalous dimension is a
consequence of the running of the strong coupling inside the
evolution kernel. Within the kinetic interpretation of parton
evolution afforded by the dynamical renormalization group, this is
a consequence of the fact recognized above at the perturbative
level that the coupling runs on scales \emph{shorter} than that of
the distribution function and hence the full running coupling
\emph{must} be included in the kernel [see the discussion after
Eq.~(\ref{newgainloss})].

In conclusion, we have given an alternative derivation of the
Altarelli-Parisi-Lipatov equation from the point of view  of the
dynamical renormalization group. While the final equation of
evolution for the parton distribution function is the same as that
obtained by Collins and Qiu~\cite{CQ}, we believe that the
derivation via the dynamical renormalization group provides
further insight into the nature of the approximation and
resummation. This approach helps clarify several important issues
in the original derivation of Collins and Qiu:
\begin{itemize}
\item[(i)]{At the level of perturbation theory, a detailed analysis
of the ``update'' equation (\ref{neweq3}) with the gain and loss
contributions given by Eq.~(\ref{newgainloss}) reveals that a
consistent treatment of the parton evolution to leading
logarithmic order should only involve the parton distribution at
the initial scale. In turn this permits to understand
unambiguously whether the strong coupling should be allowed to run
during the scale evolution. This analysis is similar to the issue
of separation of time scales in a kinetic description of
nonequilibrium phenomena. It is precisely this analysis that
ultimately leads to the correct violations of
scaling~\cite{handbook}, as a power of logarithms in
Eq.~(\ref{scal}), a consequence of the running of the coupling in
the evolution kernel.}
\item[(ii)]{The dynamical renormalization group manifestly
and systematically treats the $\ln Q^2$ corrections that arises in
the perturbative expansion as a renormalization of the parton
distribution functions and establishes a direct correspondence
between the leading logarithmic divergences arising from soft
gluon emission and the leading secular divergences in time in the
derivation of the Boltzmann equation. A common feature of such
divergences in the parton evolution equations and in quantum
kinetics is that they signal the breakdown of the perturbative
approach to the evolution dynamics for large-momentum  or
long-time   scales respectively. This breakdown of the
perturbative expansion requires a resummations which is
consistently recognized as a renormalization of the distribution
functions.}

\item[(iii)]{The dynamical renormalization
group approach reveals that the leading logarithmic approximation
in which the parton evolution equation only depends on the parton
distribution at the momentum scale at which it is probed,
corresponds to the Markovian approximation in a kinetic
description.}
\end{itemize}

We hope that the experience with non-Markovian kinetics in
nonequilibrium statistical physics can pave the way, via the
dynamical renormalization group, to understand how to consistently
treat next-to or next-next-to leading logarithmic corrections.
Beyond the leading logarithmic approximation lack of angular
ordering of gluon emission will likely result in convolutions both
in the variable $x$ and in the variable $Q^2$, i.e., a
non-Markovian parton evolution in $Q^2$ within the kinetic
interpretation. Work on these aspects is currently underway.

\bigskip

\emph{Acknowledgements.} D.B.\ and H.L.Y.\ thank J.\ Qiu for
illuminating discussions. The work of D.B.\ and S.-Y.W.\ was
supported in part by the US NSF under grants PHY-9988720,
NSF-INT-9815064 and NSF-INT-9905954. S.-Y.W.\ would like to thank
the Andrew Mellon Foundation for partial support. D.-S.L.\ was
supported by the ROC NSC through grant NSC-90-2112-M-259-010 and
NSC-90-2112-M-259-011. H.-L.Y.\ would like to thank LPTHE,
Universit\'e Pierre et Marie Curie (Paris VI) et Denis Diderot
(Paris VII) and the Department of Physics and Astronomy,
University of Pittsburgh for their hospitality. H.-L.Y.\ was
partially supported by the ROC NSC under grant
NSC-90-2112-M-001-049.

\end{document}